\documentclass[12pt,a4paper]{article}
\usepackage{epsfig}

\title{Stabilization mechanism for two-dimensional solitons in
nonlinear parametric resonance}
\author{N.V. Alexeeva \\
Department of Mathematics, University of Cape Town,\\
South Africa 
}

\begin{document}


\maketitle


\begin{abstract}
We consider a simple model system supporting stable solitons in two dimensions.
The system is the parametrically driven damped nonlinear Schr\"odinger equation,
and the soliton stabilises  for sufficiently strong damping. The
purpose of this note is to elucidate the stabilisation mechanism; we do
this by reducing the partial differential equation to a finite-dimensional
dynamical system. Our conclusion is
 that the negative feedback loop occurs via the enslaving
of the soliton's phase, locked to the driver, to its amplitude and width.

\end{abstract}

\maketitle

{\bf 1.} When a liquid layer is subjected to vertical vibration, a one-
or
two-dimensional periodic pattern forms on its surface. This phenomenon
has
been known since the celebrated Faraday
resonance experiment \cite{Muller}; more recently, it was 
found that the vertical vibration is also
 capable of sustaining 
 {\it localised\/} 2D states. These spatially localised,
temporally oscillating structures --- commonly referred to as {\it
oscillons\/}
--- were observed on the surface of granular materials
 \cite{Swinney},
Newtonian \cite{Newtonian,Astruc} and non-Newtonian \cite{nonNewtonian}
fluids.
Subsequently, stable oscillons were  reproduced in numerical simulations
within a variety of models,
including the order-parameter
equations \cite{order,Astruc}, discrete-time maps with continuous spatial
coupling \cite{map},
 semicontinuum  \cite{semicon} and hydrodynamic \cite{hydro} theories. 
 Although these simulations 
accounted for the formation of oscillons in several
particular physical
settings, they did not uncover the core of the mechanism 
which makes them immune from  the nonlinear blow-up and 
dispersive broadening. The fact that 
 stable oscillons occur in diverse physical media
and in mathematical
models of various nature,  suggests that this mechanism is simple
and 
general. It should operate whenever one has a balance 
of dispersion and nonlinearity on one hand, and of damping 
and phase-sensitive
amplification on the other.

In order to crystallise the main ingredients
of this mechanism, a simple model of  nonlinear distributed system 
exhibiting 
parametric resonance  was proposed recently \cite{us}. The model
comprises a two-dimensional lattice of diffusively coupled, vertically
vibrated,
damped 
pendula. In the present note we consider the associated amplitude equation
whose stationary soliton solutions furnish the
slowly varying amplitudes of
the lattice oscillons.
Understanding how these 2D stationary
{\it solitons\/} manage to resist the nonlinear blow-up or dispersive decay in
the amplitude equation will 
provide insights into the stabilisation of {\it oscillons\/} in vibrated media.

{\bf 2.} The amplitude equation we consider,
\begin{equation}
i \psi_t + \nabla^2 \psi+ 2|\psi|^2 \psi - \psi = h \psi^* -i \gamma
\psi,
\label{2Dnls}
\end{equation}
is the parametrically driven, damped nonlinear Schr\"odinger
(NLS) equation.
Here $\nabla^2=\partial^2 /\partial x^2 +\partial^2 /\partial y^2.$
Apart from the
pendulum lattice, eq.(\ref{2Dnls}) serves as an amplitude equation
for a wide range of nearly-conservative two-dimensional oscillatory
systems under parametric forcing.
Physically, it  
was used as a phenomenological model
of nonlinear Faraday resonance in fluids \cite{Kiyashko,Vinals,Astruc}. Independently, it
appeared in the 
context of optical parametric oscillators \cite{Sanchez}.

Two stationary
radially-symmetric soliton solutions are  given by
\begin{equation}
\psi_{\pm}= {\cal A}_{\pm} e^{- i \theta_{\pm}} \, {\cal R}({\cal A}_{\pm} r),
\label{soliton}
\end{equation}
where $r^2=x^2+y^2$;
 $${\cal A}^2_{\pm}=1 \pm  \sqrt{h^2-\gamma^2}, \quad
\theta_{+}= \frac12 \arcsin\left(\frac{\gamma}{h}\right), \quad \theta_{-}= \pi
- \frac12 \arcsin\left(\frac{\gamma}{h}\right),$$ and
${\cal R}(r)$ is the bell-shaped (monotonically decreasing) solution of
equation
\begin{equation}
 {\cal R}_{rr} +\frac 1 r {\cal R}_r- {\cal R} + 2 {\cal R}^3 =0,
  \label{master}
\end{equation}
with the boundary conditions ${\cal R}_r(0)={\cal R}(\infty)=0$.
This solution is well documented in
literature 
\cite{Rypdal}.

The soliton $\psi_+$ exists for all $h> \gamma$ while the soliton $\psi_-$
exists only in the  wedge  $\gamma<h<\sqrt{1+\gamma^2}$.
It is pertinent to add here that when $h< \gamma$,
all initial conditions are damped to zero. This follows from
the rate equation
\begin{equation}
\partial_t |\psi|^2=
2  \nabla  (|\psi|^2 \nabla \chi) +2 |\psi|^2
(h \sin 2 \chi - \gamma),
\label{rate}
\end{equation}
where $\chi$ is the phase
of the field $\psi$:  $\psi=|\psi|e^{-i \chi}$.
Defining $$N= \int |\psi|^2 d {\bf x},$$
eq.(\ref{rate}) implies $$\partial_t{N} \le 2(h-\gamma)N,$$ whence
$N(t) \to 0$  as $t \to \infty$.
We should also mention here that on the 
other side of the wedge, i.e. for $h > \sqrt{1+\gamma^2}$, 
all solutions with $\psi \to 0$ as $|x| \to \infty$ are unstable
against nonlocalised 
(continuous spectrum) perturbations \cite{BBK}.

\begin{figure}
\begin{center}
\psfig{file=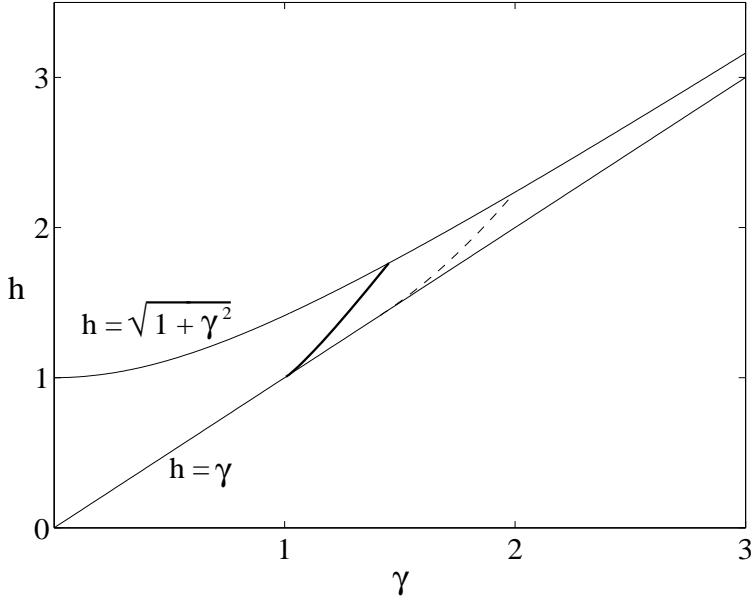,width=10cm,height=8cm}
\end{center}
\caption{\sf Stability diagram for the two-dimensional soliton
($\psi_+$).
 No stationary localised solutions
exist below the straight line $h= \gamma$,
while above the `parabola' $h=\sqrt{1+\gamma^2}$ all localised
solutions are unstable to radiation waves.  The region of
stability of the soliton   lies 
between the straight line and `parabola',
 to the right of the
 thick solid curve.
The dashed  curve
 gives the variational approximation
to the stability boundary: $h=( 1+ \gamma^4/4)^{1/2}$,
$\gamma \ge \sqrt{2}$.}
\label{chart}
\end{figure}

In the absence of the damping and driving, i.e. when $h=\gamma=0$, all localised
initial conditions in the two-dimensional NLS
equation are known to either disperse or blow-up in finite time
\cite{Rypdal,collapse}.
Recently it was shown, however, that while the soliton $\psi_-$ remains
  unstable
for all $h$ and $\gamma$, the soliton $\psi_+$ 
stabilises for sufficiently strong damping and
driving \cite{us}. 
(The smallest value of $\gamma$ for which this soliton can be stable,
is  1.006.) 
The corresponding stability chart is shown in Fig.\ref{chart}. Our
purpose is to explain, in
qualitative terms, the stabilization mechanism that is at work here.

{\bf 3.} To this end, we use the variational
approach. 
Equation (\ref{2Dnls}) is
derivable from the stationary action principle with the
Lagrangian
\begin{equation}
\label{Lagrangian_psi}
{\cal L}= e^{2 \gamma t} \, {\rm Re\/} \int  (
i \psi_t \psi^* -|\nabla \psi|^2 -|\psi|^2
+ |\psi|^4 -h \psi^2 ) d {\bf x}.
\end{equation}
Choosing a bell-shaped trial function \cite{ansatz} $$\psi=
\sqrt{A}  e^{-i \theta -(B+i \sigma) r^2},$$ 
 with $A,B,\theta,$ and $\sigma$  real functions of $t$,
the Lagrangian (\ref{Lagrangian_psi}) reduces to
\begin{equation}
{\cal L}=e^{2 \gamma t} \, \frac{A}{B}
\left[
{\dot \theta}-1  +  \frac{\dot \sigma}{2B}
- \frac{2B}{\cos^2 \phi} + \frac{A}{2} 
 - h \cos (\phi +2 \theta)
\cos \phi \right], 
\label{Lagrangian}
\end{equation}
with $\tan \phi= \sigma/B$.
The corresponding Euler-Lagrange equations are
\begin{eqnarray}
\label{parent1}
{\dot A} &=& 8 \sigma A - 2 \gamma A+ 4hA \sin(\phi+2 \theta) \cos \phi
- 2hA \sin [2(\phi+\theta)] \cos^2 \phi, 
\\
{\dot B}&=& 8 \sigma B + 2 hB \sin(\phi+ 2 \theta) \cos \phi 
-
2hB \sin [2(\phi+ \theta)] \cos^2 \phi,\\
{\dot \theta}&=& 1+ 4B- \frac32 A +  2 h\cos(\phi+ 2 \theta) \cos \phi
- h\cos [2 (\phi +  \theta)] \cos^2 \phi, \\
{\dot \sigma}&=&4 \sigma^2 -4B^2 +AB - 2hB\cos(\phi+2 \theta) \cos \phi
\nonumber
\\
& & + 2hB 
 \cos[2 (\phi+  \theta)] \cos^2 \phi. 
 \label{parent4}
\end{eqnarray}

The four-dimensional dynamical system defined
by (\ref{parent1})-(\ref{parent4}) has two fixed points
representing  the two stationary  solitons:
\begin{eqnarray*}
& & A_{\pm}= 2\left(1\pm\sqrt{h^2- \gamma^2} \right), \quad
B_{\pm}=\frac{A_{\pm}}{4}, \\ 
& & \theta_+ =
\frac{1}{2} \arcsin{\frac{\gamma}{h}}, \quad \theta_- = \pi -
\frac{1}{2} \arcsin{\frac{\gamma}{h}},\quad \sigma_{\pm}=0. 
\end{eqnarray*}
Consistently with the stability properties of the
solitons in the full PDE (\ref{2Dnls}), the fixed point 
$(A_-,\,B_-)$ is unstable for all $h$ and $\gamma$ whereas the point
 $(A_+,\, B_+)$ is unstable
 for small $\gamma$ but stabilises for larger dampings.
(More precisely,  this stationary point is stable in the region described by
$h > \sqrt{1+\gamma^4/4}$, with $\gamma \ge \sqrt{2}$ ---
see Fig.\ref{chart}.)
Therefore, the four-mode approximation captures the essentials of the
infinite-dimensional dynamics in the 
localised-waveform sector. We will now establish two constraints
reducing the number of independent degrees of freedom to two;
these constraints will eventually provide the key to 
the stability mechanism.

\begin{figure}
\begin{center}
\psfig{file=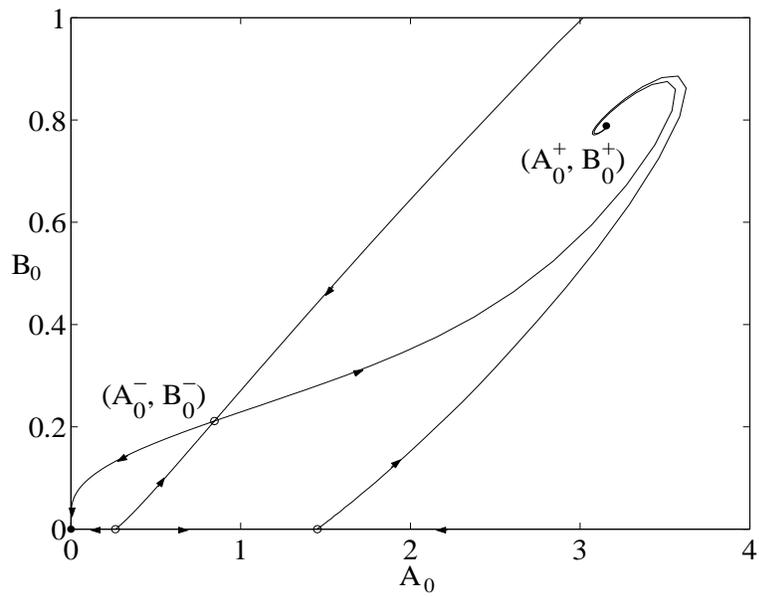,width=10cm,height=8cm}
\end{center}
\caption{\sf Phase portrait of the vector 
field (\ref{A})-(\ref{B}).
(In this plot, $c=1/3$.)
The stable manifold of the saddle $(A_0^-,B_0^-)$
separates the plane into the basins of attraction 
of the node at the origin and focus at $(A_0^+,B_0^+)$.
}
\label{fase}
\end{figure}

{\bf 4.} The two-dimensional reduction arises in the 
 overdamped limit, i.e. for large $\gamma$. In this limit,
 the dynamics should occur on a slow time-scale; 
 hence we introduce the ``slow"
time $T=t /\gamma$.
We can also expand the solution in powers of the small
parameter $\gamma^{-1}$:  
\begin{eqnarray}
A&=&A_0+\frac{1}{\gamma} A_1+..., \quad
B= B_0+ \frac{1}{\gamma} B_1 +...,\nonumber \\
\theta&=&\frac{\pi}{4} +\frac{1}{\gamma} \theta_1+..., \quad
\sigma=\frac{1}{\gamma} \sigma_1+... .\nonumber
\end{eqnarray}
Letting $h=\gamma+ c/(2 \gamma)$ with $0 \le c \le 1$,
we make sure that $h$ lies in the region of interest:
$\gamma< h< \sqrt{1+\gamma^2}$.
Substituting in (\ref{parent1})-(\ref{parent4})
and matching coefficients of like powers of
$\gamma^{-1}$, yields a  two-dimensional  system
\begin{eqnarray}
\frac{d A_0}{dT} &=& A_0[c+ 8 \sigma_1 -4\theta_1^2 + 2(\sigma_1/B_0)^2],
 \label{A} \\
 \frac{d B_0}{dT} &=& 8 \sigma_1B_0 +4 \sigma_1\theta_1+4(\sigma_1^2/B_0),
\label{B} 
\end{eqnarray}
where
\begin{eqnarray}
\label{th_si1}
\theta_1&=& \frac12+2B_0-\frac34 A_0, \\
 \sigma_1&=& \frac12 A_0B_0-2B_0^2
 \label{th_si2}.
 \end{eqnarray}
Like their parent system,
eqs.(\ref{A})-(\ref{B}) have two stationary points
in the first quadrant of the $(A_0,B_0)$-plane,  
$$B_0^{\pm}=\frac{1 \pm\sqrt{c}}{2}, \quad A_0^{\pm}=4B_0^{\pm},$$
with $\theta_1^\pm = \mp \sqrt{c}/2$ and $ \sigma_1^\pm =0$.
(Since the 
$A_0$- and $B_0$-axis are invariant manifolds, we can restrict our
attention to the first quadrant only. In particular,
fixed
points with $A_0<0$ or $B_0<0$ can have no effect on the 
dynamics in the first quadrant.) 
Like in system
(\ref{parent1})-(\ref{parent4}) with large $\gamma$, the fixed point
$(A_0^+, \, B_0^+)$ is stable (a stable focus) and the other fixed point,
$(A_0^-, \, B_0^-)$, is unstable (a saddle).
The sink at the origin is another attractor in the 
system, competing with the ``soliton" $(A_0^+,B_0^+)$. The 
corresponding basins of attraction are separated by the 
stable manifold of the saddle  (see Fig.\ref{fase}). 
As $t \to -\infty$, the top part of the separatrix
(along with many other trajectories) satisfies $A_0,B_0 \to \infty$ with
$A_0 \propto B_0^{7/2}$.

The most important conclusion of the finite-dimensional
 analysis is summarised by
eqs.(\ref{th_si1})-(\ref{th_si2}): two of the four modes are enslaved
by the other two. It is fitting 
to note here that the division into the ``masters" and ``slaves"
is somewhat arbitrary; although in (\ref{A})-(\ref{th_si2}) the amplitude
and width appear as ``master" modes and the two components
of the phase as ``slaves",  the reduction can be easily
reformulated in such a way that, for example,  
$\theta$ and $\sigma$ are the masters and $A$ and $B$ are
the slaves. All one needs to do is express $A_0$ and $B_0$ through 
$\theta_1$ and $\sigma_1$ from (\ref{th_si1}) and (\ref{th_si2}), and
substitute into
 (\ref{A}) and (\ref{B}).

{\bf 5.} 
In order to explain the stability mechanism,  we turn to 
 equation (\ref{rate}) governing the density 
 of the soliton's elementary 
 constituents, $|\psi|^2$.
 [If eq.(\ref{2Dnls}) is used to model Faraday
resonance in  granular media, 
$\int |\psi|^2$ d{\bf x} has the
meaning of the  total number of 
particles captured in
the oscillon.]
  The first term on the right-hand side of (\ref{rate}) does not
affect the total number of the  constituents.  
All it does is rearranges the constituents across the
oscillon. The second term, on the contrary, does give rise to the creation
and annihilation of particles. Since this term is proportional to 
$|\psi|^2$, the creation and annihilation occur mainly in the core of the
oscillon, where $|\psi|^2$ is not small. In the core we have $r \sim 0$
and so the creation and annihilation is controlled by $\theta$, the
uniform component of the phase
\begin{equation}
\chi=\theta+ \sigma r^2.
\label{chi} 
\end{equation}
The nonuniform part of the phase, $\sigma r^2$, 
is small in the core and plays a secondary role in
this process. Instead, the significance 
of the quantity
 $\sigma$  is in that it controls the flux of the constituents
  between the core and the periphery of
 the soliton --- see the $\nabla \chi$-term in the r.h.s. of (\ref{rate}).

 If we perturb
the stationary point $\psi_+$  in the 4-dimensional phase space of
(\ref{parent1})-(\ref{parent4}),
the variables $\theta$ and $\sigma$ will zap,
  within
a very short time $\Delta t \sim \frac{1}{\gamma}$,
onto the 2-dimensional
subspace defined by the constraints (\ref{th_si1})-(\ref{th_si2}). After this
short 
transient,
the evolution of $\theta$ and $\sigma$ will be immediately following
 that of
the soliton's amplitude
   $\sqrt{A}$ and width $1/\sqrt{B}$. Since the phase $\chi$ is 
   coupled to the driver,
 this provides a negative feedback: perturbations in $A$ and $B$
  produce only such changes in the two parts of 
  the phase  that  the new
  values of $\chi(r)$
 stimulate the recovery of the stationary values of $A$ and $B$.
 (The flat phase $\theta$ works to restore the number of constituents
 while the chirp $\sigma$ rearranges them across the soliton.)

This can be illustrated by considering $\psi$ as the envelope
of an oscillon on the surface of a granular layer, e.g.
a layer of tiny brass beads --- as in the original experiment
\cite{Swinney}.
 Imagine that we increase the amplitude $\sqrt{A}$ of the
oscillon (for example, by dropping several beads on its top): $\delta
A_0(0)>0$. Assume, for 
simplicity, that we do this without changing the oscillon's width:
$\delta B_0(0)=0$. 
From   eqs.(\ref{th_si1}),(\ref{th_si2}) it follows then that 
$\delta \theta_1(0)<0$. Since the angle $2 \chi=2(\pi/4+ \gamma^{-1}
\theta_1^+)$ is {\it acute\/} for the stationary 
point $(A_0^+,B_0^+)$ (remember, $\theta_1^+<0$ and $\sigma^+_1=0$),
the decrease in $\theta$ produces a decrease in $\sin 2 \chi$.
(Here $\chi$ is given by eq.(\ref{chi}).)
As a result,
the second term in the right-hand side of  eq.(\ref{rate}) becomes negative
which triggers the annihilation of elementary constituents. 
(In the experimental situation this simply means that the oscillon starts
``leaking"   beads to the surrounding
medium.) The annihilation continues until the original, 
stationary, value of $\chi$ (and hence,
the original value of $A=A_+$) is recovered.

Why does this mechanism not
work in the case of  the unstable fixed point,
$(A_0^-,B_0^-)$? The difference is  that in that case, $\theta_1^->0$ and so 
 $2 \chi=2(\pi/4+ \gamma^{-1} \theta_1^-)$ 
 is an {\it obtuse\/} angle. Therefore adding particles
 at the initial moment 
 of time and the resulting decrease of the phase $\chi$ 
 give rise to an {\it increase\/} of $\sin 2 \chi$.
 The second term in the right-hand side of (\ref{rate}) becomes positive and this
 triggers a further creation of elementary constituents
 (that is, more brass beads will be pulled into the
 oscillon from the surrounding layer.) Hence this time
the feedback is positive which makes the stabilisation
 impossible. 
  
 Finally, why is the large damping essential for  stability? 
  For small
 $\gamma$ the
 coupling of
 $\theta$ and $\sigma$ to $A$ and $B$ is via differential rather
 than
 algebraic equations. This time,
 the dynamics of $\theta$ and $\sigma$  is inertial
 and so the evolution of the phase  
   may not catch up with that of the
 amplitude and
 width. The feedback loop breaks down and the soliton destabilises.

In conclusion,  the stabilisation mechanism comprises two
main ingredients: (a) the enslaving of two essential degrees
of freedom (e.g. the flat and quadratic components of the 
phase) by another two (amplitude and effective width of the
soliton); and  (b)
 locking of the phase (and thereby of the amplitude
and width) to the parametric driver. 

{\bf Acknowledgment.} This project was supported by grants from the 
URC of the University of Cape Town, and
National Research Foundation of South Africa.

\end{document}